\documentstyle[amstex,amsfonts]{article}

\catcode`\@=11
\def\numberbysection{\@addtoreset{equation}{section}
        \def\theequation{\thesection.\arabic{equation}}}
\numberbysection

\begin{document}

\title{{\huge B}$_{n}^{(1)}$ {\huge and A}$_{2n}^{(2)}${\huge \ reflection
K-matrices}}
\author{{\bf \ A. Lima-Santos} \\
{\normalsize Universidade Federal de S\~{a}o Carlos, Departmento de F\'{\i}%
sica}\\
{\normalsize Caixa Postal }${\normalsize 676}${\normalsize , CEP }$%
{\normalsize 13569-905}${\normalsize , S\~{a}o Carlos, Brasil }}
\maketitle

\begin{abstract}
We investigate the regular solutions of the boundary Yang-Baxter equation
for the vertex models associated with the $B_{n}^{(1)}$ and $A_{2n}^{(2)}$
affine Lie algebras. \ In both class of models we find two general solutions
with $n+1$ free parameters. In addition, we have find $2n-1$ diagonal
solutions for $B_{n}^{(1)}$ models and $2n+1$ diagonal solutions for $%
A_{2n}^{(2)}$ models. It turns out that for each $B_{n}^{(1)}$ model there
exist a diagonal $K$-matrix with one free parameter. Moreover, a three free
parameter general solution exists for the $B_{1}^{(1)}$ model which is the
vector representation for the Zamolodchikov-Fateev model.
\end{abstract}

\vskip1truecm PACS: 75.10.-Jm; 05.90.+m

\noindent Keywords: Reflection equation; K-matrix

\section{{}Introduction}

Recently there have been many efforts to introduce boundaries into
integrable systems for possible applications in condensed matter physics and
statistical systems with non-periodic boundary conditions. The bulk
Boltzmann weights of an exactly solvable lattice system are usually the
non-zero matrix elements of a ${\cal R}$-matrix ${\cal R}(u)$ which
satisfies the Yang-Baxter equation \cite{Baxter, KIB, ABR}. 
\begin{equation}
{\cal R}_{12}(u){\cal R}_{13}(u+v){\cal R}_{23}(v)={\cal R}_{23}(v){\cal R}%
_{13}(u+v){\cal R}_{12}(u),  \label{int.1}
\end{equation}%
in $V^{1}\otimes V^{2}\otimes V^{3}$, where ${\cal R}_{12}={\cal R}\otimes 
{\bf 1}$, ${\cal R}_{23}={\bf 1}\otimes {\cal R}$, etc.

An ${\cal R}$ matrix is said to be regular if it satisfies the property $%
{\cal R}(0)=P$, where $P$ is the permutation matrix in $V^{1}\otimes V^{2}$: 
$P(\left| \alpha \right\rangle \otimes \left| \beta \right\rangle )=\left|
\beta \right\rangle \otimes \left| \alpha \right\rangle $ for $\left| \alpha
\right\rangle ,\left| \beta \right\rangle \in V$. \ In addition, we will
require \cite{MN1} that ${\cal R}(u)$ satisfies the following properties%
\begin{eqnarray}
{\rm regularity} &:&{\cal R}_{12}(0)=f(0)^{1/2}P_{12},  \nonumber \\
{\rm unitarity} &:&{\cal R}_{12}(u){\cal R}_{12}^{t_{1}t_{2}}(-u)=f(u), 
\nonumber \\
{\rm PT-symmetry} &:&P_{12}{\cal R}_{12}(u)P_{12}={\cal R}%
_{12}^{t_{1}t_{2}}(u),  \nonumber \\
{\rm cros}\text{sin}{\rm g-symmetry} &:&{\cal R}_{12}(u)=U_{1}{\cal R}%
_{12}^{t_{2}}(-u-\rho )U_{1}^{-1},  \label{int.2}
\end{eqnarray}%
where $f(u)$ is some scalar function, $t_{i}$ denotes transposition in the
space $i$ , $\rho $ is the crossing parameter and $U$ a crossing matrix,
both being specific to each model \cite{Bazha}. Note that unitarity and
crossing-symmetry together imply the useful relation%
\begin{equation}
M_{1}{\cal R}_{12}^{t_{2}}(-u-\rho )M_{1}^{-1}{\cal R}_{12}^{t_{1}}(u-\rho
)=f(u).  \label{int.3}
\end{equation}%
where%
\begin{equation}
M=U^{t}U=M^{t}.  \label{int.4}
\end{equation}

The boundaries entail new physical quantities called reflection matrices
which depend on the boundary properties. The boundary weights then follow
from $K$-matrices which satisfy boundary versions of the Yang-Baxter
equation \cite{MN1, Skly, Chere1}: the reflection equation%
\begin{equation}
{\cal R}_{12}(u-v)K_{1}^{-}(u){\cal R}%
_{12}^{t_{1}t_{2}}(u+v)K_{2}^{-}(v)=K_{2}^{-}(v){\cal R}%
_{12}(u+v)K_{1}^{-}(u){\cal R}_{12}^{t_{1}t_{2}}(u-v),  \label{int.5}
\end{equation}%
and the dual reflection equation%
\[
{\cal R}_{12}(-u+v)(K_{1}^{+})^{t_{1}}(u)M_{1}^{-1}{\cal R}%
_{12}^{t_{1}t_{2}}(-u-v-2\rho )M_{1}(K_{2}^{+})^{t_{2}}(v) 
\]%
\begin{equation}
=(K_{2}^{+})^{t_{2}}(v)M_{1}{\cal R}_{12}(-u-v-2\rho
)M_{1}^{-1}(K_{1}^{+})^{t_{1}}(u){\cal R}_{12}^{t_{1}t_{2}}(-u+v).
\label{int.6}
\end{equation}%
In this case there is an isomorphism \ between $K^{-}$ and $K^{+}$ :%
\begin{equation}
K^{-}(u):\rightarrow K^{+}(u)=K^{-}(-u-\rho )^{t}M.  \label{int.7}
\end{equation}%
Therefore, given a solution to the reflection equation (\ref{int.5}) we can
also find a solution to the dual reflection equation (\ref{int.6}).

Due to the significance of the $K$-matrices in the construction of
integrable models with open boundaries, a lot of work has been directed to
the study \cite{Chere2, Vega, MN2, Inami} and classification \cite{Batch,
Lima1, Liu} of \ them.

While the investigation of particular solutions has been made to a number of
lattice models, Batchelor {\it at al} \cite{Batch} have derived diagonal
solutions for face and vertex models associated with several affine Lie
algebras, \ the classification of all possible $K$-matrices has been a
harder problem. However, recently we have proposed a procedure which allows
the classification of the $D_{n+1}^{(2)}$ \cite{MG, Lima2} and $%
A_{n-1}^{(1)} $ reflection $K$-matrices \cite{Abad, Lima3}. \ In spite of
these papers we decided to continue in this line in order to include the $%
B_{n}^{(1)}$ and $A_{2n}^{(2)}$ regular reflection $K$-matrices.

We have organized this paper as follows. In Section $2$ we choose the
reflection equations and in Section $3$ their general solutions are derived.
In Section $4$ reduced solutions are presented. The last section is reserved
for the conclusion. The $K$-matrices for $B_{1}^{(1)}$ and $A_{2}^{(2)}$
models are written explicitly in appendices.

\section{B$_{n}^{(1)}$ and A$_{2n}^{(2)}$ reflection equations}

The ${\cal R}$-matrix for the vertex models associated to the $B_{n}^{(1)}$
and $A_{2n}^{(2)}$ affine Lie algebras as presented by Jimbo in \cite{Jimbo}
has the form 
\begin{eqnarray}
{\cal R} &=&a_{1}\sum_{i\neq i^{\prime }}E_{ii}\otimes
E_{ii}+a_{2}\sum_{i\neq j,j^{\prime }}E_{ii}\otimes
E_{jj}+a_{3}\sum_{i<j,i\neq j^{\prime }}E_{ij}\otimes
E_{ji}+a_{4}\sum_{i>j,i\neq j^{\prime }}E_{ij}\otimes E_{ji}  \nonumber \\
&&+\sum_{i,j}a_{ij}\ E_{ij}\otimes E_{i^{\prime }j^{\prime }}  \label{re.1}
\end{eqnarray}%
where $E_{ij}$ denotes the elementary $2n+1$ by $2n+1$ matrices ($%
(E_{ij})_{ab}=\delta _{ia}\delta _{ib}$) and the Boltzmann weights with
functional dependence on the spectral parameter $u$ are given by 
\begin{eqnarray}
a_{1}(u) &=&({\rm e}^{u}-q^{2})({\rm e}^{u}-\zeta ),\qquad \ a_{2}(u)=q({\rm %
e}^{2u}-1)({\rm e}^{2u}-\zeta ),  \nonumber \\
a_{3}(u) &=&-(q^{2}-1)({\rm e}^{u}-\zeta ),\qquad \ a_{4}(u)={\rm e}%
^{u}a_{3}(u)  \label{re.2}
\end{eqnarray}%
and%
\begin{equation}
a_{ij}(u)=\left\{ 
\begin{array}{cc}
(q^{2}{\rm e}^{u}-\zeta )({\rm e}^{u}-1) & (i=j,i\neq i^{^{\prime }}) \\ 
q({\rm e}^{u}-\zeta )({\rm e}^{u}-1)+(\zeta -1)(q^{2}-1){\rm e}^{u} & 
(i=j,i=i^{^{\prime }}) \\ 
(q^{2}-1)\ \left( \zeta q^{\overset{\_}{i}-\overset{\_}{j}}({\rm e}%
^{u}-1)-\delta _{ij^{\prime }}({\rm e}^{u}-\zeta )\right) & (i<j) \\ 
(q^{2}-1){\rm e}^{u}\left( q^{\overset{\_}{i}-\overset{\_}{j}}({\rm e}%
^{u}-1)-\delta _{ij^{\prime }}({\rm e}^{u}-\zeta )\right) & (i>j)%
\end{array}%
\right.  \label{re.3}
\end{equation}%
where $q=e^{-2\eta }$ denotes an arbitrary parameter and $\overset{\_}{i}$
and $\ i^{\prime }$ are defined by

\begin{equation}
\overset{\_}{i}=\left\{ 
\begin{array}{c}
i+1/2\ \qquad \quad \ \ \ (i<n+1) \\ 
i\quad \qquad \qquad \ \ \ \ \ \ (i=n+1) \\ 
i-1/2\ \qquad \quad \ \ \ (i>n+1)%
\end{array}%
\right. \qquad {\rm and}\qquad \ i^{\prime }=2n+2-i.  \label{re.4}
\end{equation}%
Here $\zeta =q^{2n-1}$ for the $B_{n}^{(1)}$ models and $\zeta =-q^{2n+1}$
for the $A_{2n}^{(2)}$ models.

Regular solutions of the reflection equation (\ref{int.5}) mean that the
matrix $K^{-}(u)$ in the form 
\begin{equation}
K_{-}(u)=\sum_{i,j=1}^{2n+1}k_{i,j}(u)\ E_{ij}  \label{re.5}
\end{equation}%
satisfies the condition 
\begin{equation}
k_{i,j}(0)=\delta _{i,j},\quad \qquad i,j=1,2,...,2n+1.  \label{re.6}
\end{equation}

Substituting (\ref{re.1}) and (\ref{re.5}) into (\ref{int.5}), we will get $%
(2n+1)^{4}$ functional equations for the $k_{i,j}$ matrix elements, many of
which are dependent. In order to solve them, we shall proceed in the
following way. First we consider the $(i,j)$ component of the matrix
equation (\ref{int.5}). By differentiating it with respect to $v$ and taking 
$v=0$, we get algebraic equations involving the single variable $u$ and $%
(2n+1)^{2}$ parameters 
\begin{equation}
\beta _{i,j}=\frac{dk_{i,j}(v)}{dv}|_{v=0}\qquad i,j=1,2,...,2n+1
\label{re.7}
\end{equation}%
Second, these equations are denoted by $E[i,j]=0$ and collected into blocks $%
B[i,j]$ , $i=1,...,2n(n+1)+1$ and $j=i,i+1,...,2n(n+1)+1-i$, defined by 
\begin{equation}
B[i,j]=\left\{ 
\begin{array}{c}
E[i,j]=0,\  \\ 
E[j,i]=0,\  \\ 
E[(2n+1)^{2}+1-i,(2n+1)^{2}+1-j]=0, \\ 
E[(2n+1)^{2}+1-j,(2n+1)^{2}+1-i]=0.%
\end{array}%
\right. \   \label{re.8}
\end{equation}%
For a given block $B[i,j]$, the equation $E[(2n+1)^{2}+1-i,(2n+1)^{2}+1-j]=0$
can be obtained from the equation $E[i,j]=0$ by interchanging 
\begin{equation}
k_{i,j}\longleftrightarrow k_{i^{\prime },j^{\prime }},\quad \beta
_{i,j}\longleftrightarrow \beta _{i^{\prime },,j^{\prime }},\quad
a_{3}\longleftrightarrow a_{4},\quad \ a_{ij}\leftrightarrow a_{i^{\prime
}j^{\prime },}.  \label{re.9}
\end{equation}%
and the equation $E[j,i]=0$ is obtained from the equation $E[i,j]=0$ by the
interchanging 
\begin{equation}
k_{ij}\longleftrightarrow k_{ji},\quad \beta _{ij}\longleftrightarrow \beta
_{ji},\quad a_{ij}\longleftrightarrow a_{j^{\prime }i^{\prime }}\ .\quad
\label{re.11}
\end{equation}%
In that way, we will have a better control about these equations.

Since the ${\cal R}$-matrix (\ref{re.1}) satisfies unitarity, {\small PT}
invariances and crossing symmetry, the matrix $K^{+}(u)$\ is obtained using 
\begin{equation}
K^{+}(u)=K^{-}(-u-2\rho )^{{\rm t}}M.  \label{re.12}
\end{equation}%
where from \cite{Bazha} we have%
\begin{equation}
M={\rm diag}(1,q,\ldots ,q^{n-1},1,q^{1-n},q^{2-n},\ldots ,q^{-1},1),\quad
\rho =(2n-1)\eta ,  \label{re.13}
\end{equation}%
for the $B_{n}^{(1)}$ models and%
\begin{equation}
M={\rm diag}(1,q,\ldots ,q^{n-1},1,q^{1-n},q^{2-n},\ldots ,q^{-1},1),\quad
\rho =-2(2n+1)\eta -i\pi  \label{re.14}
\end{equation}%
for the $A_{2n}^{(2)}$ models.

Here we observe that the cases $n=1$ are well known: $B_{1}^{(1)}$ is the
Zamolodchikov-Fateev model \cite{Zamo1} or the $A_{1}^{(1)}$ model (spin-$1$
representation) for which $M=1$ and $\rho =\eta $ while $A_{2}^{(2)}$ is the
Izergin-Korepin model \cite{IK} which has $M={\rm diag}({\rm e}^{2\eta },1,%
{\rm e}^{-2\eta })$ and $\rho =-6\eta -i\pi $.

Having built a common ground for $B_{n}^{(1)}$ and $A_{2n}^{(2)}$ models, we
may proceed in order to find their reflection solutions.

\section{General solutions}

Analyzing the reflection equations (\ref{int.5})\ for $B_{n}^{(1)}$ and $%
A_{2n}^{(2)}$ models we can see that the simplest are those involving only
two matrix elements of the type $k_{i,i^{^{\prime }}}$\ (secondary
diagonal). They belong to the blocks $B[1,2n+3],B[1,4n+5],B[1,6n+7],...$,
and we chose to express their solutions in terms of the element $k_{1,2n+1}$%
with $\beta _{1,2n+1}\neq 0$:%
\begin{equation}
k_{i,i^{^{\prime }}}=\left( \frac{\beta _{i,i^{^{\prime }}}}{\beta _{1,2n+1}}%
\right) k_{1,2n+1}.  \label{gs.1}
\end{equation}%
Next, we look at the last blocks of the collection $\{B[1,j]\}$. Here we can
write the matrix elements of the first row $k_{1,j}\ (j\neq 1,2n+1)$ in
terms of the element $k_{1,2n+1}$ and their transpose in terms of the
element $k_{2n+1,1}$. From the last blocks of the collection $\{B[2n+3,j]\}$%
, the matrix elements of the second row $k_{2,j}\ (j\neq 2,2n)$ are
expressed in terms of $k_{2,2n}$ \ and their transpose in terms of $k_{2n,2}$%
. Following this procedure with the collections $\{B[4n+5,j]\},\{B[6n+7,j]%
\},...,$ we will be able to write all non-diagonal matrix elements as:%
\begin{equation}
k_{i,j}=\left( \frac{a_{1}a_{11}-a_{2}^{2}}{%
a_{3}a_{4}a_{11}^{2}-a_{2}^{2}a_{12}a_{21}}\right) \left( \beta
_{i,j}a_{3}a_{11}-\beta _{j^{^{\prime }},i^{^{\prime }}}a_{2}a_{ij^{^{\prime
}}}\right) \frac{k_{1,2n+1}}{\beta _{1,2n+1}}\qquad (j<i^{^{\prime }})
\label{gs.2}
\end{equation}%
and%
\begin{equation}
k_{i,j}=\left( \frac{a_{1}a_{11}-a_{2}^{2}}{%
a_{3}a_{4}a_{11}^{2}-a_{2}^{2}a_{12}a_{21}}\right) \left( \beta
_{i,j}a_{4}a_{11}-\beta _{j^{^{\prime }},i^{^{\prime }}}a_{2}a_{ij^{^{\prime
}}}\right) \frac{k_{1,2n+1}}{\beta _{1,2n+1}}\qquad (j>i^{^{\prime }})
\label{gs.3}
\end{equation}%
where we have used \ (\ref{gs.1}) and the identities%
\begin{equation}
a_{ij}=a_{j^{^{\prime }}i^{^{\prime }}}\qquad {\rm and}\qquad
a_{1j}a_{j1}=a_{12}a_{21}\qquad (j\neq 1).  \label{gs.4}
\end{equation}

Taking into account the Boltzmann weights (\ref{re.2}) and (\ref{re.3}), we
substitute these expressions\ in the remaining \ reflection equations and
look at for those without diagonal matrix elements $k_{i,i}$, in order to
fix some parameters $\beta _{i,j}$ ($i\neq j$). For instance, from the
diagonal blocks $B[i,i]$ one can see that their equations are solved by the
relations%
\begin{equation}
\beta _{i,j}k_{j,i}=\beta _{j,i}k_{i,j}  \label{gs.5}
\end{equation}%
\ provided that%
\begin{equation}
\beta _{i,j}\beta _{j^{^{\prime }},i^{^{\prime }}}=\beta _{j,i}\beta
_{i^{^{\prime }},j^{^{\prime }}}.  \label{gs.6}
\end{equation}

This procedure is simple but too long due to a large number of equations
with non-diagonal terms. After some algebraic manipulations we found two
possibilities to express the parameters for the matrix elements below the
secondary diagonal ($\beta _{i,j}$ with $j>i^{^{\prime }}$) in terms of
those above of the secondary diagonal: 
\begin{equation}
\beta _{i,j}=\left\{ 
\begin{array}{c}
\epsilon q^{(\overset{\_}{i}-\overset{\_}{i^{^{\prime }}})/2+j-2n-1}\ \beta
_{j^{^{\prime }},i^{^{\prime }}}\qquad {\rm for}\qquad j>n+1 \\ 
\\ 
\epsilon q^{(\overset{\_}{j}-\overset{\_}{j^{^{\prime }}})/2+i-2n-1}\ \beta
_{j^{^{\prime }},i^{^{\prime }}}\qquad {\rm for}\qquad j\leq n+1%
\end{array}%
\right.  \label{gs.7}
\end{equation}%
where $\epsilon =\pm 1$ for $B_{n}^{(1)}$ models and $\epsilon =\pm \frac{i}{%
q}$ ($i=\sqrt{-1}$) for $A_{2n}^{(2)}$ models.

These relations simplify considerably the expressions for the non-diagonal
matrix elements (\ref{gs.2}) and (\ref{gs.3}):%
\begin{equation}
k_{i,j}(u)=\left\{ 
\begin{array}{c}
\beta _{i,j}{\cal G}_{n}^{(\epsilon )}(u),\qquad \qquad \quad \ \ \ \ \ \ \
\ (j<i^{^{\prime }}) \\ 
\\ 
\beta _{i,i^{^{\prime }}}\left( \frac{q^{n-3/2}+\epsilon {\rm e}^{u}}{%
q^{n-3/2}+\epsilon }\right) {\cal G}_{n}^{(\epsilon )}(u)\qquad
(j=i^{^{\prime }}) \\ 
\\ 
\beta _{i,j}{\rm e}^{u}{\cal G}_{n}^{(\epsilon )}(u)\qquad \qquad \ \ \ \ \
\ \ \ \ \ (j>i^{^{\prime }})%
\end{array}%
\right.  \label{gs.8}
\end{equation}%
where ${\cal G}_{n}^{(\epsilon )}(u)$ is defined conveniently by a
normalization of $k_{1,2n+1}(u):$ 
\begin{equation}
{\cal G}_{n}^{(\epsilon )}(u)=\frac{1}{\beta _{1,2n+1}}\left( \frac{%
q^{n-3/2}+\epsilon }{q^{n-3/2}+\epsilon {\rm e}^{u}}\right) k_{1,2n+1}(u).
\label{gs.9}
\end{equation}

{\small \ Now, }we substitute these expressions in the remaining equations
and look at the equations of the type 
\begin{equation}
F(u){\cal G}_{n}^{(\epsilon )}(u)=0  \label{gs.10}
\end{equation}%
where $F(u)=\sum_{k}f_{k}(\left\{ \beta _{ij}\right\} ){\rm e}^{ku}$. \ The
constraint equations $f_{k}(\left\{ \beta _{ij}\right\} )\equiv 0,\forall k$%
, can be solved in terms of the $2n+1$ parameters. Of course the expressions
for $k_{ij}$ will depend on our choice of these parameter. After some
attempts we concluded the choice $\beta _{12}$, $\beta _{13}$, ..., $\beta
_{1,2n+1}$ and $\beta _{21}$ as the most appropriate for our purpose.

Taking into account all fixed\ parameters in terms of these $2n+1$
parameters, we can rewrite the $k_{ij}$ $(i\neq j)$ matrix elements for $%
n\neq 1$ in the following way:

The matrix elements in the secondary diagonal of $K^{-}(u)$ are given by 
\begin{equation}
k_{i,i^{\prime }}(u)=\frac{q^{\overset{\_}{i}-2n+3/2}}{(q+1)^{2}}\frac{\beta
_{1,i^{\prime }}^{2}}{\beta _{1,2n+1}}\left( q^{n-3/2}+\epsilon \right)
\left( q^{n-3/2}+\epsilon {\rm e}^{u}\right) {\cal G}_{n}^{(\epsilon
)}(u),\qquad (i\neq 1,2n+1)  \label{gs.11}
\end{equation}%
and 
\begin{equation}
k_{2n+1,1}(u)=q^{2n-3}\left( \frac{\beta _{21}}{\beta _{1,2n}}\right)
^{2}\beta _{1,2n+1}\left( \frac{q^{n-3/2}+\epsilon {\rm e}^{u}}{%
q^{n-3/2}+\epsilon }\right) {\cal G}_{n}^{(\epsilon )}(u).  \label{gs.12}
\end{equation}%
For the first row and the first column of $K^{-}(u)$ the matrix elements are 
\begin{eqnarray}
k_{1,j}(u) &=&\beta _{1,j}\ {\cal G}_{n}^{(\epsilon )}(u),\qquad \ \ \ \ \ \
\ \ \ \ \ \ \ \ \ \ \ \ \ \ (j\neq 1,2n+1), \\
&&  \nonumber \\
k_{i,1}(u) &=&q^{\overset{\_}{i}-5/2}\left( \frac{\beta _{21}\beta
_{1,i^{^{\prime }}}}{\beta _{1,2n}}\right) {\cal G}_{n}^{(\epsilon
)}(u),\qquad (i\neq 1,2n+1)  \label{gs13.b}
\end{eqnarray}%
and in the last column and the last row, we have%
\begin{eqnarray}
k_{i,2n+1}(u) &=&\epsilon q^{\overset{\_}{i}-n-1}\beta _{1,i^{\prime }}{\rm e%
}^{u}{\cal G}_{n}^{(\epsilon )}(u),\qquad \ \ \ \ \ \ \ \ \ \ \ (i\neq
1,2n+1), \\
&&  \nonumber \\
k_{2n+1,j}(u) &=&\epsilon q^{n-3/2}\left( \frac{\beta _{21}\beta _{1,j}}{%
\beta _{1,2n}}\right) {\rm e}^{u}{\cal G}_{n}^{(\epsilon )}(u),\qquad (j\neq
1,2n+1).  \label{gs.14b}
\end{eqnarray}%
The remaining non-diagonal matrix elements are given by%
\begin{equation}
k_{ij}(u)=\left\{ 
\begin{array}{c}
q^{\overset{\_}{i}-n}\left( \frac{q^{n-3/2}+\epsilon }{q+1}\right) \left( 
\frac{\beta _{1,i^{\prime }}\beta _{1,j}}{\beta _{1,2n+1}}\right) {\cal G}%
_{n}^{(\epsilon )}(u)\qquad \ \ \ \ \ \ \ \ \ (j<i^{\prime }) \\ 
\\ 
\epsilon q^{\overset{\_}{i}-2n+1/2}\left( \frac{q^{n-3/2}+\epsilon }{q+1}%
\right) \left( \frac{\beta _{1,i^{\prime }}\beta _{1,j}}{\beta _{1,2n+1}}%
\right) {\rm e}^{u}{\cal G}_{n}^{(\epsilon )}(u)\quad \ \ \ (j>i^{\prime })%
\end{array}%
\right. .  \label{gs.15}
\end{equation}

At this point the remaining reflection equations involve the diagonal matrix
elements $k_{ii}(u)$ and $4n+2$ parameters. However, almost all of them have
only two diagonal matrix elements. Working with those containing consecutive
elements we can get the following relations:

\begin{equation}
k_{i+1,i+1}(u)=\left\{ 
\begin{array}{c}
k_{i,i}(u)+\left( \beta _{i+1,i+1}-\beta _{i,i}\right) {\cal G}%
_{n}^{(\epsilon )}(u)\qquad \ \ \ \ \ \ \ \ \ \ \ \ \ \ \ \ \ \ (1\leq i\leq
n) \\ 
\\ 
k_{i,i}(u)+\left( \beta _{i+1,i+1}-\beta _{i,i}\right) {\rm e}^{u}{\cal G}%
_{n}^{(\epsilon )}(u)\qquad (n+2<i\leq 2n+1)%
\end{array}%
\right.  \label{gs.16}
\end{equation}%
and two special relations%
\begin{eqnarray}
k_{n+1,n+1}(u) &=&k_{n,n}(u)+\left( \beta _{n+1,n+1}-\beta _{n,n}\right) 
{\cal G}_{n}^{(\epsilon )}(u)-\epsilon {\cal F}_{n}^{(\epsilon )}(u), \\
&&  \nonumber \\
k_{n+2,n+2}(u) &=&k_{n+1,n+1}(u)+\left( \beta _{n+2,n+2}-\beta
_{n+1,n+1}\right) {\rm e}^{u}{\cal G}_{n}^{(\epsilon )}(u)-q^{n-1/2}{\cal F}%
_{n}^{(\epsilon )}(u),  \label{gs.17}
\end{eqnarray}%
where we have defined a scalar function%
\begin{equation}
{\cal F}_{n}^{(\epsilon )}(u)=\left( \frac{q^{2-n}}{q+1}\right) \left( \frac{%
q^{n-3/2}+\epsilon }{q+1}\right) \left( \frac{\beta _{1,n}\beta _{1,n+2}}{%
\beta _{1,2n+1}}\right) \left( {\rm e}^{u}-1\right) {\cal G}_{n}^{(\epsilon
)}(u).  \label{gs.18}
\end{equation}%
It means that we can express all diagonal matrix elements in terms of \ $%
{\cal G}_{n}^{(\epsilon )}(u)$ and $k_{11}(u)$:

\begin{equation}
k_{i,i}(u)=k_{11}(u)+(\beta _{i,i}-\beta _{11}){\cal G}_{n}^{(\epsilon )}(u)
\label{gs.19}
\end{equation}%
for $1\leq i\leq n,$%
\begin{equation}
k_{n+1,n+1}(u)=k_{11}(u)+(\beta _{n+1,n+1}-\beta _{11}){\cal G}%
_{n}^{(\epsilon )}(u)-\epsilon {\cal F}_{n}^{(\epsilon )}(u)  \label{gs.20}
\end{equation}%
and%
\begin{equation}
k_{i,i}(u)=k_{11}(u)+(\beta _{n+1,n+1}-\beta _{11}){\cal G}_{n}^{(\epsilon
)}(u)+(\beta _{i,i}-\beta _{n,n}){\rm e}^{u}{\cal G}_{n}^{(\epsilon
)}(u)-\left( q^{n-1/2}+\epsilon \right) {\cal F}_{n}^{(\epsilon )}(u)
\label{gs.21}
\end{equation}%
for $n+2\leq i\leq 2n+1.$

The most important fact is that these recurrence relations are closed by the
solution of the block $B[2n+2,4n+2]$. \ From this block we can get another
expression for $k_{2n+1,2n+1}$: 
\begin{equation}
k_{2n+1,2n+1}(u)={\rm e}^{2u}k_{11}(u)+(\beta _{2n+1,2n+1}-\beta _{11}-2)%
{\rm e}^{u}\left( \frac{q^{n-3/2}+\epsilon {\rm e}^{u}}{q^{n-3/2}+\epsilon }%
\right) {\cal G}_{n}^{(\epsilon )}(u).  \label{gs.22}
\end{equation}%
Taking $i=2n+1$ into (\ref{gs.21}) and comparing with (\ref{gs.22}) we can
find $k_{11}(u)$ without solve any additional equation:

\begin{eqnarray}
k_{11}(u) &=&\left( \frac{2{\rm e}^{u}-\left( \beta _{n+1,n+1}-\beta
_{11}\right) ({\rm e}^{u}-1)}{{\rm e}^{2u}-1}\right) {\cal G}_{n}^{(\epsilon
)}(u)-\left( \frac{q^{n-1/2}+\epsilon }{{\rm e}^{2u}-1}\right) {\cal F}%
_{n}^{(\epsilon )}(u)  \nonumber \\
&&-\epsilon \left( \frac{\beta _{2n+1,2n+1}-\beta _{11}-2}{%
q^{n-3/2}+\epsilon }\right) \frac{{\rm e}^{u}}{{\rm e}^{u}+1}{\cal G}%
_{n}^{(\epsilon )}(u)  \label{gs.23}
\end{eqnarray}%
Here we recall that $\epsilon =\pm 1$ for the $B_{n}^{(1)}$ models and $%
\epsilon =\pm \frac{i}{q}$ for the $A_{2n}^{(2)}$ models.

These relations were derived for $n>1$. It turns out that the cases $n=1$
are ruled out and their general solution are presented in appendices.
Indeed, these relations hold for the $A_{2}^{(2)}$ model after some
modifications.

Before we substitute these expressions into the reflection equations we
first need fix some parameters. To do this we can, for instance, look at the
combination ${\rm e}^{u}k_{11}(u)+k_{22}(u)$ from several block equations.
Consistency conditions of these results will give us all constraint
equations to fasten the $4n+2$ remaining parameters. Following this
procedure we can fix $2n-1$ diagonal parameters, $\beta _{i,i}$ ($i\neq
1,2n+1$):

\begin{equation}
\beta _{i,i}=\left\{ 
\begin{array}{c}
\beta _{11}+(-1)^{n}\left( \frac{q^{n-3/2}+\epsilon }{q^{n-3/2}}\right)
\left( \sum_{j=0}^{i-2}(-q)^{j}\right) \frac{\beta _{1,n}\beta _{1,n+2}}{%
\beta _{1,2n+1}}\ \ \ \ \ \ \ \ \ \ \ \ \ \ \ \ \ (1<i\leq n) \\ 
\\ 
\beta _{n+2,n+2}-\epsilon \sqrt{q}\left( \frac{q^{n-3/2}+\epsilon }{q^{n-3/2}%
}\right) \left( \sum_{j=0}^{i-n-3}(-q)^{j}\right) \frac{\beta _{1,n}\beta
_{1,n+2}}{\beta _{1,2n+1}}\quad \ \ \ (n+2<i<2n+1)%
\end{array}%
\right.  \label{gs.24}
\end{equation}%
where%
\begin{eqnarray}
\beta _{n+1,n+1} &=&\beta _{11}+\frac{q(q^{n-3/2}+\epsilon )}{q+1}\left\{ 
\frac{\beta _{1.n+1}^{2}}{\beta _{1,2n+1}}+\left( \frac{q^{n}+(-1)^{n}(q+1)-%
\epsilon \sqrt{q}}{q^{n-1/2}(q+1)}\right) \frac{\beta _{1,n}\beta _{1,n+2}}{%
\beta _{1,2n+1}}\right\}  \label{gs.25} \\
&&  \nonumber \\
\beta _{n+2,n+2} &=&\beta _{11}+\frac{(q^{n-3/2}+\epsilon
)(q^{n-1/2}-\epsilon )}{q^{n-3/2}(q+1)}\left\{ \frac{\beta _{1.n+1}^{2}}{%
\beta _{1,2n+1}}\right.  \nonumber \\
&&+\left. \left( \frac{2q^{n}+(-1)^{n}(1+q)+\epsilon \sqrt{q}(q-1)}{%
(q^{n-1/2}-\epsilon )(q+1)}\right) \frac{\beta _{1,n}\beta _{1,n+2}}{\beta
_{1,2n+1}}\right\}  \label{gs.26}
\end{eqnarray}%
and $n-1$ non-diagonal parameters 
\begin{eqnarray}
\beta _{21} &=&-\frac{1}{q^{2n-4}}\left( \frac{q^{n-3/2}+\epsilon }{q+1}%
\right) ^{2}\frac{\beta _{12}\beta _{1,2n}^{2}}{\beta _{1,2n+1}^{2}}
\label{gs.27} \\
&&  \nonumber \\
\beta _{1,j} &=&(-1)^{n+j}\frac{\beta _{1,n}\beta _{1,n+2}}{\beta _{1,2n+2-j}%
},\qquad j=2,3,...,n-1  \label{gs.28}
\end{eqnarray}%
Next, we can substitute these expression into the block $B[2n+1,2n+2]$ to
fix the two last parameters $\beta _{2n+1,2n+1}$ and $\beta _{1,n}$: 
\begin{eqnarray}
\beta _{2n+1,2n+1} &=&\beta _{11}+2+(-1)^{n}\frac{\epsilon (\epsilon ^{2}-1)%
}{q^{n-9/2}}\left( \frac{q^{n-3/2}+\epsilon }{q+1}\right) ^{2}\frac{\beta
_{1,n}\beta _{1,n+2}}{\beta _{1,2n+1}},  \label{gs.29} \\
&&  \nonumber \\
\beta _{1,n} &=&(-1)^{n}\frac{\epsilon (q+1)}{\left( \epsilon \sqrt{q}%
-(-1)^{n}\right) ^{2}}\left( \frac{\beta _{1,n+1}^{2}}{\beta _{1,n+2}}-2%
\frac{q^{n-3/2}(q+1)}{(q^{n-3/2}+\epsilon )(q^{n-1/2}-\epsilon )}\frac{\beta
_{1,2n+1}}{\beta _{1,n+2}}\right) .  \label{gs.30}
\end{eqnarray}%
In that way, we have derived the $B_{n}^{(1)}$ and $A_{2n}^{(2)}$ general
solutions for the boundary Yang-Baxter equations (\ref{int.5}) and their
dual equations (\ref{int.6}), using (\ref{re.12}).

The final result are two solutions with $n+2$ parameters $\beta
_{1,n+1},\beta _{1,n+2},\ldots ,\beta _{1,2n+1}$ and $\beta _{11}$. The
number of free parameters is $n+1$ because we still have to use the regular
condition (\ref{re.6}). For instance, we can choose the arbitrary functions
as%
\begin{equation}
k_{1,2n+1}(u)=\frac{1}{2}\beta _{1,2n+1}({\rm e}^{2u}-1)  \label{gs.31}
\end{equation}%
and fix the parameter $\beta _{11}$ by the regular condition.

Let us summarize these results: First, we have from (\ref{gs.11}) to (\ref%
{gs.15}) all non-diagonal matrix elements. Second, the diagonal matrix
elements are obtained using (\ref{gs.19}), (\ref{gs.20}) and (\ref{gs.21})
with $k_{11}(u)$ given by (\ref{gs.23}). Finally we substitute into these
matrix elements all fixed parameters which are given by (\ref{gs.24})--(\ref%
{gs.30}). The final expressions for these matrix elements are very
cumbersome.

\section{Reduced solutions}

In the previous section we have used the condition $\beta _{1,2n+1}\neq 0$
and considered all $\beta _{i,j}$ different from zero. Nevertheless, the
cases $\beta _{1,2n+1}=0$ and all $\beta _{i,j}=0$ $(j\neq i,i^{\prime })$
should be considered separately.\ 

In this section we will relax these conditions. First we observe that the
parameters $\beta _{i,j}$ and $\beta _{j,i}$ $(i\neq j)$ are linked by the
relation (\ref{gs.5}) and the constraint equations (\ref{gs.6}) and (\ref%
{gs.7}) together the normalization condition (\ref{re.6}) imply that%
\begin{equation}
{\rm if\ \ }\beta _{i,j}=0,\ \ {\rm then\ }\left\{ 
\begin{array}{c}
k_{ij}(u)=0,\quad {\rm for}\quad i\neq j \\ 
{\rm or} \\ 
k_{ij}(u)=1,\quad {\rm for}\quad i=j%
\end{array}%
\right.  \label{red.1}
\end{equation}%
In our general solutions we have $n+1$ free parameters. Therefore,we can
generate several reduced solutions taking one or more free parameters equal
to zero.

From the reflection equations for $n>1$ one can see that the vanishing of an 
$k_{i,i^{\prime }}$ element implies that the only elements different from
zero are those in the $i$th-row and in the $i^{\prime }$th-column of $K^{-}$%
, {\it i.e.} 
\begin{equation}
\beta _{i,i^{\prime }}=0\Rightarrow \left\{ k_{i,l}(u)=0\ (l\neq i)\quad 
{\rm and}\quad k_{l,i^{\prime }}(u)=0\ (l\neq i^{\prime })\right\}
\label{red.2}
\end{equation}%
Similar consideration holds for their tranpose elements. This procedure
gives us several $K$-matrices which can be obtained from the general
solution after an appropriate choice of parameters. For example, to get
diagonal solutions from the general solutions it is enough to take the limit 
$\beta _{i,i^{\prime }}\rightarrow 0$. This limit procedure is not trivial
due to our initial choice for the free parameters and one can lose
solutions. We can outline this problem solving the reflection equations
again but now with diagonal $K$-matrices.

\subsection{Diagonal solutions}

Solving the reflection equation we found $2n-1$ diagonal solutions for the $%
B_{n}^{(1)}$ models and $2n+1$ diagonal solutions for the $A_{2n}^{(2)}$
models. We denoted these solutions by ${\Bbb K}_{\epsilon }^{[p]}$\ and $%
{\Bbb K}_{[\beta ]}$. The matrix elements of ${\Bbb K}_{\epsilon }^{[p]}$
are given by%
\begin{eqnarray}
k_{11}(u) &=&k_{22}(u)=\cdots =k_{p,p}(u)={\rm e}^{-u}  \nonumber \\
k_{p+1,p+1}(u) &=&k_{p+1,p+1}(u)=\cdots =k_{2n+1-p,2n+1-p}(u)=\frac{%
q^{2p-n-1/2}{\rm e}^{u}+\epsilon }{q^{2p-n-1/2}+\epsilon {\rm e}^{u}} 
\nonumber \\
k_{2n+2-p,2n+2-p}(u) &=&k_{2n+3-p,2n+3-p}(u)=\cdots =k_{2n+1,2n+1}(u)={\rm e}%
^{u}  \label{red.3}
\end{eqnarray}%
while the matrix elements of ${\Bbb K}_{[\beta ]}$ are%
\begin{eqnarray}
k_{11}(u) &=&\left( \frac{\beta ({\rm e}^{-u}-1)+2}{\beta ({\rm e}^{u}-1)+2}%
\right)  \nonumber \\
k_{22}(u) &=&\cdots =k_{n+1,n+1}(u)=\cdots =k_{2n,2n}(u)=1  \nonumber \\
k_{2n+1,2n+1}(u) &=&\left( \frac{\beta (q^{2n-3}{\rm e}^{u}-1)+2}{\beta
(q^{2n-3}{\rm e}^{-u}-1)+2}\right)  \label{red.4}
\end{eqnarray}%
where $\beta =\beta _{n+1,n+1}-\beta _{11}$ is the free parameter. These
solutions hold for $n\geq 1$.

For the $B_{n}^{(1)}$ models the diagonal solutions are ${\Bbb K}_{\epsilon
}^{[p]}\ (p=2,3,...,n)$ with $\epsilon =\pm 1$ and the one parameter
solution ${\Bbb K}_{[\beta ]}$. For the $A_{2n}^{(2)}$ models the diagonal
solutions are ${\Bbb K}_{\epsilon }^{[p]}\ (p=1,2,...,n)$ with $\epsilon
=\pm \frac{i}{q}$ and the trivial solution which is multiple of the identity.

Here we notice that the solutions ${\Bbb K}_{\epsilon }^{[p=n]}$ were
already computed by Batchelor {\it at al} \cite{Batch}. Moreover, the cases $%
n=1$ are well known: The diagonal solution for $B_{1}^{(1)}$ model is the
matrix ${\Bbb K}_{[\beta ]}$ (\ref{red.4})%
\begin{equation}
{\Bbb K}_{[\beta ]}=\left( 
\begin{array}{ccc}
\frac{\beta ({\rm e}^{-u}-1)+2}{\beta ({\rm e}^{u}-1)+2} & 0 & 0 \\ 
0 & 1 & 0 \\ 
0 & 0 & \frac{\beta (q^{-1}{\rm e}^{u}-1)+2}{\beta (q^{-1}{\rm e}^{-u}-1)+2}%
\end{array}%
\right)  \label{red.5}
\end{equation}%
while for the $A_{2}^{(2)}$ model the diagonal solutions are%
\begin{equation}
{\Bbb K}^{[0]}=\left( 
\begin{array}{ccc}
1 & 0 & 0 \\ 
0 & 1 & 0 \\ 
0 & 0 & 1%
\end{array}%
\right) ,\quad {\Bbb K}_{i/q}^{[1]}=\left( 
\begin{array}{ccc}
{\rm e}^{-u} & 0 & 0 \\ 
0 & \frac{q\sqrt{q}{\rm e}^{u}+i}{q\sqrt{q}+i{\rm e}^{u}} & 0 \\ 
0 & 0 & {\rm e}^{u}%
\end{array}%
\right) ,\quad {\Bbb K}_{-i/q}^{[1]}=\left( 
\begin{array}{ccc}
{\rm e}^{-u} & 0 & 0 \\ 
0 & \frac{q\sqrt{q}{\rm e}^{u}-i}{q\sqrt{q}-i{\rm e}^{u}} & 0 \\ 
0 & 0 & {\rm e}^{u}%
\end{array}%
\right) .  \label{red.6}
\end{equation}%
These solutions were obtained for the first time by Menzincescu and
Nepomechie in \cite{MN2} and in \cite{MN3}, together with Rittenberg.

\subsection{The case $\protect\beta _{i,j}=0$ $(j\neq i,i^{\prime })$}

Following our classification procedure we still have to consider the case
for which all $\beta _{i,j}=0$ ($j\neq i,i^{\prime }$). The corresponding
non-zero matrix elements are $k_{i,i}$ (main diagonal) and $k_{i,i^{\prime
}} $ (secondary diagonal).

For the $B_{n}^{(1)}$ models it is not a new type of solution because it is
a limit of the general solutions. Taking the limit $\beta _{i,j}\rightarrow
0 $ ($j\neq i,i^{\prime }$) into the general solutions for each $B_{n}^{(1)}$
model ($n\geq 1$) we can find one reduced solution whose normalized matrix
elements are given by%
\begin{eqnarray}
k_{11}(u) &=&k_{22}(u)=\cdots =k_{n,n}(u)=1  \nonumber \\
k_{n+1,n+1}(u) &=&\frac{{\rm e}^{2u}-q}{1-q}  \nonumber \\
k_{n+2,n+2}(u) &=&k_{n+3,n+3}(u)=\cdots =k_{2n+1,2n+1}(u)={\rm e}^{2u}
\label{case.1}
\end{eqnarray}%
and%
\begin{equation}
k_{i,i^{\prime }}(u)=\left\{ 
\begin{array}{c}
\frac{1}{2}\beta _{i,i^{\prime }}({\rm e}^{2u}-1)\qquad i<n+1 \\ 
\\ 
\frac{q}{(q-1)^{2}}\frac{2}{\beta _{i^{\prime },i}}\left( {\rm e}%
^{2u}-1\right) \qquad i>n+1%
\end{array}%
\right.  \label{case.2}
\end{equation}%
Observe that this limit procedure reduced to $n$ the number of free
parameters.

For each $A_{2n}^{(2)}$ model ($n\geq 1$) we find the following $K$-matrix%
\begin{eqnarray}
k_{11}(u) &=&k_{22}(u)=\cdots =k_{n,n}(u)=1+\beta _{11}({\rm e}^{u}-1) 
\nonumber \\
k_{n+1,n+1}(u) &=&\beta _{11}{\rm e}^{u}-\frac{{\rm e}^{2u}-q}{1-q}(\beta
_{11}-1)  \nonumber \\
k_{n+2,n+2}(u) &=&k_{n+3,n+3}(u)=\cdots =k_{2n+1,2n+1}(u)={\rm e}^{2u}\left[
1+\beta _{11}({\rm e}^{-u}-1)\right]  \label{case.3}
\end{eqnarray}%
and%
\begin{equation}
k_{i,i^{\prime }}(u)=\left\{ 
\begin{array}{c}
\frac{1}{2}\beta _{i,i^{\prime }}({\rm e}^{2u}-1)\qquad i<n+1 \\ 
\\ 
q\left( \frac{\beta _{11}-1}{q-1}\right) ^{2}\frac{2}{\beta _{i^{\prime },i}}%
\left( {\rm e}^{2u}-1\right) \qquad i>n+1%
\end{array}%
\right.  \label{case.4}
\end{equation}%
Note that in this solution, the number of free parameters is same of the
general solutions. Therefore it is not a limit of our general solution. This
is a new type of $K$-matrix with $n+1$ free parameters. In particular, for
the Izergin-Korepin model, it has the form%
\begin{equation}
K^{-}=\left( 
\begin{array}{ccc}
1+\beta _{11}({\rm e}^{u}-1) & 0 & \frac{1}{2}\beta _{13}({\rm e}^{2u}-1) \\ 
0 & \beta _{11}{\rm e}^{u}-\frac{{\rm e}^{2u}-q}{1-q}(\beta _{11}-1) & 0 \\ 
\frac{1}{\beta _{13}}\frac{2q(\beta _{11}-1)^{2}}{(q-1)^{2}}({\rm e}^{2u}-1)
& 0 & {\rm e}^{2u}\left[ 1+\beta _{11}({\rm e}^{-u}-1)\right]%
\end{array}%
\right)  \label{case.5}
\end{equation}%
This two free parameter solution for $A_{2}^{(2)}$ model was derived for the
first time by Kim \cite{Kim}.

\section{Conclusion}

We still are believing that a direct computation should be a starting point
to obtain and classify the solutions of the boundary Yang-Baxter equations.

After a systematic study of their functional equations we find solutions for
the vertex models associated with the $B_{n}^{(1)}$ and $A_{2n}^{(2)}$
affine Lie algebras. For the $B_{n}^{(1)}$ models there is a ruled out
solution when $n=1$. But, for $n>1$ their solutions following the same rules
used for obtain the $A_{2n}^{(2)}$ solutions. These models have a rich
spectrum of diagonal solutions: For each $B_{n}^{(1)}$ model there are $2n-1$
diagonal solutions, being $2n-2$ without free parameters and one $1$-free
parameter solution. For the $A_{2n}^{(2)}$ models we found $2n+1$ diagonal
solutions without any free parameters.

To complete the classification for all non-exceptional Lie algebras we still
have to consider the vertex models associated with the $C_{n}^{(1)}$, $%
D_{n}^{(1)}$,and $A_{2n-1}^{(2)}$ Lie algebras.

\bigskip {\bf Acknowledgment:} This work was supported in part by Funda\c{c}%
\~{a}o de Amparo \`{a} Pesquisa do Estado de S\~{a}o Paulo--FAPESP--Brasil
and by Conselho Nacional de Desenvol\-{}vimento--CNPq--Brasil.

\bigskip \appendix

\section{The B$_{1}^{(1)}$ general solution}

In this appendix we will present the $K$-matrix solutions for the $%
B_{1}^{(1)}$ model. The corresponding $K^{-}$ matrix has the form%
\begin{equation}
K^{-}=\left( 
\begin{array}{ccc}
k_{11} & k_{12} & k_{13} \\ 
k_{21} & k_{22} & k_{23} \\ 
k_{31} & k_{32} & k_{33}%
\end{array}%
\right) .  \label{b1.1}
\end{equation}%
Analyzing the reflection equations we can see that the relations (\ref{gs.7}%
) vanish. Therefore we do not have the simplified form for non-diagonal
elements (\ref{gs.8}) in terms of the function ${\cal G}(u)$.

Solving the reflection equations we find the following non-diagonal elements%
\begin{eqnarray}
k_{21}(u) &=&\frac{\beta _{21}}{\beta _{12}}k_{12}(u),\quad k_{12}(u)=\left( 
\frac{\sqrt{q}\beta _{23}({\rm e}^{u}-1)+\beta _{12}(q{\rm e}^{u}-1)}{q{\rm e%
}^{2u}-1}\right) \frac{k_{13}(u)}{\beta _{13}},  \nonumber \\
k_{32}(u) &=&\frac{\beta _{21}}{\beta _{12}}k_{23}(u),\quad k_{23}(u)=\left( 
\frac{\sqrt{q}\beta _{12}({\rm e}^{u}-1)+\beta _{23}(q{\rm e}^{u}-1)}{q{\rm e%
}^{2u}-1}\right) \frac{{\rm e}^{u}k_{13}(u)}{\beta _{13}},  \nonumber \\
k_{31}(u) &=&\left( \frac{\beta _{21}}{\beta _{12}}\right) ^{2}k_{13}(u)
\label{b1.2}
\end{eqnarray}%
where%
\begin{equation}
\beta _{21}=-\sqrt{q}\left( \frac{(q-1)\beta _{12}\beta _{23}-2(q+1)\beta
_{13}}{q^{2}-1}\right) \frac{\beta _{12}}{\beta _{13}^{2}}  \label{b1.3}
\end{equation}%
In the same way, we also could not find a recurrence relation of the type (%
\ref{gs.16}) for the diagonal elements. Instead of that, we find the
diagonal matrix elements by a direct computation:

\begin{eqnarray}
k_{11}(u) &=&2\frac{k_{13}(u)}{\beta _{13}({\rm e}^{2u}-1)}  \nonumber \\
&&-\left( \frac{\sqrt{q}\left[ (q{\rm e}^{u}-1)\beta _{12}^{2}+({\rm e}%
^{u}-q)\beta _{23}^{2}\right] +(q+1)(q{\rm e}^{u}-1)\beta _{12}\beta _{23}}{%
(1+q)(q{\rm e}^{2u}-1)({\rm e}^{u}+1)}\right) \frac{k_{13}(u)}{\beta
_{13}^{2}}  \nonumber \\
k_{22}(u) &=&-2\frac{({\rm e}^{2u}-q)k_{13}(u)}{\beta _{13}(q-1)({\rm e}%
^{2u}-1)}  \nonumber \\
&&+\frac{1}{(1+q)(q{\rm e}^{2u}-1)({\rm e}^{u}+1)}\left\{ \sqrt{q}{\rm e}%
^{u}(q{\rm e}^{u}-1)(\beta _{12}^{2}+\beta _{23}^{2})\right.  \nonumber \\
&&+\left. \left[ ({\rm e}^{u}+q)(q{\rm e}^{2u}-1)+2q{\rm e}^{u}({\rm e}%
^{u}-1)\right] \beta _{12}\beta _{23}\right\} \frac{k_{13}(u)}{\beta
_{13}^{2}}  \nonumber \\
k_{33}(u) &=&2\frac{{\rm e}^{2u}k_{13}(u)}{\beta _{13}({\rm e}^{2u}-1)} 
\nonumber \\
&&-\left( \frac{\sqrt{q}\left[ (q{\rm e}^{u}-1)\beta _{23}^{2}+({\rm e}%
^{u}-q)\beta _{12}^{2}\right] +(q+1)(q{\rm e}^{u}-1)\beta _{12}\beta _{23}}{%
(1+q)(q{\rm e}^{2u}-1)({\rm e}^{u}+1)}\right) \frac{{\rm e}^{2u}k_{13}(u)}{%
\beta _{13}^{2}}  \label{b1.4}
\end{eqnarray}%
This is the $3$-parameter solution for the $B_{1}^{(1)}$ model or the
Zamolodchikov-Fateev model, which was obtained for the first time by Inami 
{\it at al} \cite{Inami}.

Note that in the limit $\beta _{23}=\pm \beta _{12}$ $\Longleftrightarrow $ $%
\beta _{32}=\pm \beta _{21}$\ this solution is unfolded in two $2$-parameter
solutions which satisfy, up to minor modifications, the procedure used in
the previous section to find the $B_{n}^{(1)}$ $(n>1)$ solutions. The
corresponding diagonal solution is given by (\ref{red.5}).

\section{The A$_{2}^{(2)}$ general solutions}

In this appendix we consider the $K$-matrix solution for the Izergin-Korepin
model and try to understand the words ''up to minor modifications'' used
previously.

Solving the $A_{2}^{(2)}$ reflection equations we still have the relations (%
\ref{gs.7}):%
\begin{equation}
\beta _{23}=\epsilon \beta _{12},\quad \beta _{32}=\epsilon \beta _{21}
\label{a2.1}
\end{equation}%
where $\epsilon =\pm i/q$. Let us to consider the case $\epsilon =i/q.$

The non-diagonal matrix elements can be read from (\ref{gs.8})

\begin{eqnarray}
k_{12}(u) &=&\beta _{12}{\cal G}(u),\quad k_{21}(u)=\beta _{21}{\cal G}%
(u),\quad k_{23}(u)=\beta _{23}{\rm e}^{u}{\cal G}(u)  \nonumber \\
k_{31}(u) &=&\beta _{31}{\cal G}(u),\quad k_{32}(u)=\beta _{32}{\cal G}(u)
\label{a2.2}
\end{eqnarray}%
where we have used the notation ${\cal G}(u)$ for the function ${\cal G}%
_{1}^{(\epsilon )}(u)$%
\begin{equation}
{\cal G}(u)=\frac{1}{\beta _{13}}\left( \frac{\sqrt{q}+i}{\sqrt{q}+i{\rm e}%
^{u}}\right) k_{13}(u)  \label{a2.3}
\end{equation}%
and $\beta _{31}$ is the last non-diagonal parameter fixed before we
consider the diagonal term 
\begin{equation}
\beta _{31}=\left( \frac{\beta _{21}}{\beta _{12}}\right) ^{2}\beta _{13}.
\label{a2.4}
\end{equation}

For the diagonal terms we still have the special recurrence relations (\ref%
{gs.17})%
\begin{eqnarray}
k_{22}(u) &=&k_{11}(u)+(\beta _{22}-\beta _{11}){\cal G}(u)-\frac{i}{q}{\cal %
P}(u),  \nonumber \\
k_{33}(u) &=&k_{22}(u)+(\beta _{33}-\beta _{22}){\rm e}^{u}{\cal G}(u)-\sqrt{%
q}{\cal P}(u),  \label{a2.5}
\end{eqnarray}%
but with a new scalar function ${\cal P}(u)$ different from ${\cal F}%
_{1}^{(\epsilon )}$ given by (\ref{gs.18})%
\begin{equation}
{\cal P}(u)=\frac{\sqrt{q}\beta _{21}\beta _{13}}{\beta _{12}}\left( \frac{%
{\rm e}^{u}-1}{\sqrt{q}+i}\right) {\cal G}(u).  \label{a2.6}
\end{equation}%
Again, the block $B[4,6]$ close the recurrence relation%
\begin{equation}
k_{33}(u)={\rm e}^{2u}k_{11}(u)+(\beta _{33}-\beta _{11}-2){\rm e}^{u}{\cal G%
}(u)\left( \frac{\sqrt{q}+i{\rm e}^{u}}{\sqrt{q}+i}\right) ,  \label{a2.7}
\end{equation}%
which gives us the $k_{11}(u)$ element%
\begin{eqnarray}
k_{11}(u) &=&\left( \frac{2{\rm e}^{u}-\left( \beta _{22}-\beta _{11}\right)
({\rm e}^{u}-1)}{{\rm e}^{2u}-1}\right) {\cal G}(u)-\left( \frac{\sqrt{q}+%
\frac{i}{q}}{{\rm e}^{2u}-1}\right) {\cal P}(u)  \nonumber \\
&&-i\left( \frac{\beta _{33}-\beta _{11}-2}{\sqrt{q}+i}\right) \frac{{\rm e}%
^{u}}{{\rm e}^{u}+1}{\cal G}(u)  \label{a2.8}
\end{eqnarray}%
and we are leaving with the problem to find $\beta _{33}$, $\beta _{22}$ and 
$\beta _{21}$%
\begin{eqnarray}
\beta _{33} &=&\beta _{11}+2+i\left( \frac{q^{2}+1}{q}\right) \frac{\beta
_{21}\beta _{13}}{\beta _{12}},  \nonumber \\
\beta _{22} &=&\beta _{11}+\left( \frac{2q^{3/2}}{q^{3/2}-i}\right) -\left( 
\frac{1+q-iq^{3/2}}{\sqrt{q}}\right) \frac{\beta _{21}\beta _{13}}{\beta
_{12}}  \label{a2.9}
\end{eqnarray}%
and%
\begin{equation}
\beta _{21}=\left( \frac{2iq}{(\sqrt{q}+i)(q^{3/2}-i)}\right) \frac{\beta
_{12}}{\beta _{13}}-\left( \frac{i}{\sqrt{q}(q+1)}\right) \frac{\beta
_{12}^{2}}{\beta _{13}^{2}}.  \label{a2.10}
\end{equation}

Differences from the $A_{2n}^{(2)}$ ($n>1$) to $A_{2}^{(2)}$ procedure are
due to our previous choice of the free parameters. It means that we can not
take the limit $n\rightarrow 1$ in the $A_{2n}^{(2)}$ general solutions to
get the solutions for the $A_{2}^{(2)}$ model.

Finally, we observe that there is an apparent simplification in these
calculus when we are comparing with those presented in \cite{Lima1}.
However, after we substitute the fixed parameters the final\ form for this
solution is still cumbersome. Nevertheless, there is an equivalent solution
for the $A_{2}^{(2)}$ model derived by Nepomechie \cite{Nepo} which looks
simpler.

There are three diagonal solutions for this model which were already
presented in (\ref{red.6}) as well as a second type of solution obtained by
Kim was presented in (\ref{case.5}).

\end{document}